\documentclass[prb,twocolumn,showpacs,superscriptaddress]{revtex4}%
\usepackage{amsfonts}
\usepackage{amsmath}
\usepackage{amssymb}
\usepackage{longtable}
\usepackage{supertabular}
\usepackage{graphicx}%
\def\beq{\begin{equation}}
\def\eeq{\end{equation}}

\providecommand{\U}[1]{\protect\rule{.1in}{.1in}}

\begin{document}
\title{Quantum molecular dynamics simulations of thermophysical properties of fluid ethane}
\author{Yujuan Zhang}
\affiliation{LCP, Institute of Applied Physics and Computational
Mathematics, P.O. Box 8009, Beijing 100088, People's Republic of
China}
\author{Cong Wang}
\affiliation{LCP, Institute of Applied Physics and Computational
Mathematics, P.O. Box 8009, Beijing 100088, People's Republic of
China}\affiliation{Center for Applied Physics and Technology, Peking
University, Beijing 100871, People's Republic of China}
\author{Fawei Zheng}
\affiliation{LCP, Institute of Applied Physics and Computational
Mathematics, P.O. Box 8009, Beijing 100088, People's Republic of
China}
\author{Ping Zhang}
\thanks{Corresponding author. Electronic mail: zhang\_ping@iapcm.ac.cn}
\affiliation{LCP, Institute of Applied Physics and Computational
Mathematics, P.O. Box 8009, Beijing 100088, People's Republic of
China} \affiliation{Center for Applied Physics and Technology,
Peking University, Beijing 100871, People's Republic of China}

\begin{abstract}
We have performed first-principles molecular-dynamics simulations
based on density-functional theory to study the thermophysical
properties of ethane under extreme conditions. We present new
results for the equation of state of fluid ethane in the warm dense
region. The optical conductivity is calculated via the
Kubo-Greenwood formula from which the dc conductivity and optical
reflectivity are derived. The close correlation between the
nonmetal-metal transition of ethane and its decomposition, that
ethane dissociates significantly into molecular and/or atomic
hydrogen and some long alkane chains, has been systematically
studied by analyzing the optical conductivity spectra, pair
correlation functions, electronic density of states and charge
density distribution of fluid ethane.
\end{abstract}
 \maketitle

\section{INTRODUCTION}
Pressure-induced transformations of hydrocarbons under extreme
conditions have recently drawn extensive attention and have made
giant achievement in astrophysics \cite{Bene1999}. Specially, the
planetary model with non-axisymmetric magnetic fields generated
within a thin fluid shell has successfully revealed the
non-axisymmetric nature of the Uranian and Neptunian magnetic fields
\cite{Stan2006,Stan2004}. As the interiors of these planets are
composed of the fluids containing species, such as, C, H, N, O,
\emph{etc}, the knowledge of physical properties of fluid
hydrocarbons (e.g., methane, ethane and benzene) under extreme
conditions, such as equation of states (EOS), electrical
conductivity and structure, is important in this context.

Saturated hydrocarbon ethane (C$_2$H$_6$) has been reported as one
of the most important decomposition products of the giant planets
\cite{Rich,Li2011,Gao2010}. For instance, Richters \emph{et al.}
\cite{Rich} have reported that ethane is formed in the decomposition
process of methane at high temperature ($T$=4000 K) and high
pressure ($P$$\approx$100 GPa). The observation of atmospheric
ethane in Neptune, where it might be brought up from the deep
interior by convection process, could be explained by the formation
of ethane. Quantum molecular dynamics (QMD) simulations of Li
\emph{et al.} \cite{Li2011} also indicated that methane, which is a
major constitute of the \textquotedblleft ice\textquotedblright\
layer in Uranus and Neptune, mainly converts into ethane at density
of 1.5 g/cm$^3$ along the principal Hugoniot. Meanwhile, ethane is
an important fluid industrially, and the second member of the
vitally interesting alkane series. Its chemical kinetics and
thermophysical properties are of great interest for gas-turbine
engines, high-speed propulsion, and materials synthesis
\cite{Vrie2007}. Pyrolysis and oxidation of ethane have been studied
behind reflected shocked waves \cite{Hida2000}. The EOS of ethane
with temperatures ranging from 90 K to 625 K and pressures less than
70 MPa has been obtained by using Benedict-Webb-Rubin equation
\cite{Frie1991}. However, such studies on ethane are limited to
narrow temperature and pressure region. More importantly, the
nonmetal-metal transition (NMT) of fluid hydrocarbons is a crucial
and hot topic recently. For instance, methane and benzene have been
reported to transform to metallic fluids under extreme conditions,
where the NMT are associated with the rapid C-H bond breaking and
the formation of atomic or molecular hydrogen
\cite{Li2011,Wang2010}. The NMT of the system has also been found to
significantly affect the optical properties, such as the reflectance
and absorption coefficient. The optical reflectivity of fluid
deuterium has been successfully detected in dynamic compression
experiments \cite{Cell2000}. To date, the thermophysical properties
of ethane under extreme conditions have not been reported yet.
Considering the present facts that (i) ethane also denotes a typical
prototype and plays an important role in hydrocarbons, (ii) ethane
is one of the most important decomposition products in the
extreme-condition matter evolution process of the giant planets, and
(iii) as a same kind of alkane species, the differences and
relations of the chemical dissociation properties between methane
and ethane are interesting, it is thus highly necessary to study the
equation of state, chemical decomposition and recomposition picture,
electronic and optical properties of fluid ethane under extreme
conditions by using an efficient method.

Materials under extreme conditions are complex, where partial
dissociation, ionization and electronic degeneration coexist. QMD
simulations, based on finite temperature density functional theory
(FT-DFT) \cite{Leno2000}, offer a powerful tool to study
thermophysical properties of complex materials under such extreme
conditions \cite{Coll1995,Hohl1993,Kres2010,Desj2002}. Combined with
the Kubo-Greenwood formulation \cite{Kubo1957,Gree1958}, the
electrical and optical properties of various systems can be
accurately calculated \cite{Maze2003,Maze2005}.

In the present work, we have performed QMD simulations to study the
thermophysical properties of fluid ethane. By analyzing pair
correlation functions, atomic structure, and charge density
distribution, we have predicted the chemical picture of the shocked
ethane. In particular, the NMT for fluid ethane can be directly
associated to its dissociation, which is analogous to that of
molecular fluid methane. Additionally, the optical properties of
fluid ethane are also discussed in the chemical picture.

\section{COMPUTATIONAL METHODS}
\subsection{First-principles molecular dynamics}
In this study, we have performed simulations for ethane by employing
the Vienna \emph{ab} \emph{initio} simulation package (VASP)
plane-wave pseudopotential code, which was developed at the
Technical University of Vienna \cite{Kres1993,Kres1996}. Electrons
are fully quantum mechanically treated through solving the Kohn-Sham
equations for a set of orbitals and energies within a plane-wave
FT-DFT formulation \cite{Leno2000}, where the electronic states are
populated according to the Fermi-Dirac distribution at temperature
$T_e$. The ion-electron interactions are represented by a projector
augmented wave pseudopotential. The exchange-correlation functional
is determined by generalized gradient approximation with the
parametrization of Perdew-Burke-Ernzerhof \cite{Perd1996}. Atoms
move classically according to the forces originated from the
interactions of ions and electrons. The isokinetic ensemble (NVT) is
employed by No$\acute{s}$e-Hoover thermostat \cite{Nose1984} and the
system is controlled in local thermodynamical equilibrium by setting
the electron temperature $T_e$ and the ion temperature $T_i$ to be
equal. Pseudopotentials with cutoff radius of 1.10 Bohr for carbon
and hydrogen are adopted, and 650 eV plane-wave cutoff energy is
employed in the simulations.

In all the simulations, a total number of 128 atoms (16 C$_2$H$_6$
molecules) are included in a supercell with periodic boundary
condition. We perform finite-temperature, fixed-volume molecular
dynamics simulations for selected densities ranging from 0.7 to
2.8 g/cm$^3$ along 3000, 4000 and 8000 K isotherms. For molecular
dynamic simulations, only $\Gamma$ point of the Brillouin zone is
included. All the dynamic simulations last 10 ps with time steps
of 0.1-1.0 fs according to different conditions. For each pressure
and temperature, the system is equilibrated within 1-2 ps. The
isotherm equation of states data and pair correlation functions
are obtained by averaging over the final 5 ps simulations for all
the particles.
\subsection{Optical and electronic properties}
The electronic properties of dense ethane along isotherms are
calculated in a spatial configuration of all atoms at a single time
step within a MD trajectory. For dispelling the correlation time,
the electronic conductivity are evaluated for 15-20 selected
snapshots taken from the QMD trajectory. The selected snapshots or
configurations are spaced by at least the correlation time. A
4$\times$4$\times$4 Monkhorst-Pack $k$ point mesh \cite{Monk1976}
has been used to calculate the electronic structure. The dynamic
conductivity is derived from the Kubo-Greenwood formula as follows
\cite{Kubo1957,Gree1958}:
\begin{eqnarray}
\sigma_{1}(\omega) & = & \frac{2\pi}{3\omega\Omega}\underset{\textbf{k}}{\sum}W(\textbf{k})\overset{N}{\underset{j=1}{\sum}}\overset{N}{\underset{i=1}{\sum}}\overset{3}{\underset{\alpha=1}{\sum}}[f(\epsilon_{i},\textbf{k})-f(\epsilon_{j},\textbf{k})]\nonumber \\
 &  & \times|\langle \Psi_{j,\textbf{k}}|\nabla_{\alpha}|\Psi_{i,\textbf{k}}\rangle |^{2}\delta(\epsilon_{j,\textbf{k}}-\epsilon_{i,\textbf{k}}-\hbar\omega)
\;\;,
\end{eqnarray}
where $\Psi_{i,\textbf{k}}$ is Kohn-Sham eigenstate, with
corresponding to eigenvalue $\epsilon_{i,\textbf{k}}$, and
occupation number $f(\epsilon_{i},\mathbf{k})$. $\Omega$ is the
volume of the supercell. W(\textbf{k}) is the K-point weighting
factor. The $i$ and $j$ summations range over $N$ discrete bands
included in the calculation, and $\alpha$ over the three spatial
directions.

The imaginary part of conductivity $\sigma_{2}$ is obtained via the Kramers-Kronig relation as
\begin{eqnarray}
\sigma_{2}(\omega) & = &
-\frac{2}{\pi}P\int\frac{\sigma_{1}(\nu)\omega}{(\nu^{2}-\omega^{2})}d\nu,
\end{eqnarray}
where $P$ is the principal value of the integral and $\nu$ is
frequency. The complex dielectric function $\epsilon$ follows immediately from the two parts of the conductivity by the following equations
\begin{eqnarray}
\epsilon_{1}(\omega) & = & 1-\frac{4\pi}{\omega}\sigma_{2}(\omega);
\epsilon_{2}(\omega)=\frac{4\pi}{\omega}\sigma_{1}(\omega).
\end{eqnarray}
The real part $n(\omega)$ and the imaginary part $k(\omega)$ of the refraction index have the following relation with the dielectric function
\begin{eqnarray}
\epsilon (\omega)= \epsilon_{1}(\omega)+i\epsilon_{2}(\omega) & = &
[n(\omega)+ik(\omega)]^2.
\end{eqnarray}
From these quantities, the absorption coefficient $\alpha(\omega)$ and reflectivity $r(\omega)$ are derived
\begin{eqnarray}
\alpha(\omega) & = & \frac{4\pi}{n(\omega)c}\sigma_1(\omega),
\end{eqnarray}
\begin{eqnarray}
r(\omega) & = &
\frac{[1-n(\omega)]^{2}+k(\omega)^{2}}{[1+n(\omega)]^{2}+k(\omega)^{2}}.
\end{eqnarray}

\section{RESULTS AND DISCUSSIONS}
\subsection{Equation of state and structure}

\begin{center}
\tablefirsthead{%
\topcaption{The simulated data of the density, pressure, and the
standard deviation of pressure.} \hline \hline
$\rho$ (g/cm$^3$)&$T$ (K)&$P$ (GPa)&$\Delta$$P$ (GPa)\\
\hline}
\tablehead{%
\multicolumn{4}{l}{\small\sl (continued)}\\
\hline \hline$\rho$ (g/cm$^3$)&$T$ (K)&$P$ (GPa)&$\Delta$$P$ (GPa)\\
\hline}
\tabletail{%
\hline} \tablelasttail{\hline\hline}

\tablefirsthead{%
\hline \hline
$\rho$ (g/cm$^3$)&$T$ (K)&$P$ (GPa)&$\Delta$$P$ (GPa)\\
\hline}
\tablehead{%
\multicolumn{4}{l}{\small\sl (continued)}\\
\hline \hline
$\rho$ (g/cm$^3$)&$T$ (K)&$P$ (GPa)&$\Delta$$P$ (GPa)\\
\hline}
\tabletail{%
\hline} \tablelasttail{\hline\hline}

\tablefirsthead{%
\hline \hline
$\rho$ (g/cm$^3$)&$T$ (K)&$P$ (GPa)&$\Delta$$P$ (GPa)\\
\hline}
\tablehead{%
\multicolumn{4}{l}{\small\sl (continued)}\\
\hline \hline
$\rho$ (g/cm$^3$)&$T$ (K)&$P$ (GPa)&$\Delta$$P$ (GPa)\\
\hline}
\tabletail{%
\hline} \tablelasttail{\hline\hline} \topcaption{The simulated data
of the density, pressure, and the standard deviation of pressure.}
\begin{supertabular}{r@{\hspace{8mm}}r@{\hspace{8mm}}r@{\hspace{8mm}}r@{\hspace{8mm}}}
0.80&3000&7.94&1.04\\
0.90&3000&11.24&2.05\\
1.00&3000&14.57&2.21\\
1.20&3000&24.84&2.54\\
1.40&3000&38.72&2.99\\
1.60&3000&57.66&3.49\\
1.80&3000&81.86&3.91\\
2.00&3000&110.38&4.20\\
2.30&3000&165.22&5.60\\
2.50&3000&209.59&5.58\\
2.80&3000&280.04&6.72\\
0.80&4000&8.54&1.07\\
0.90&4000&11.90&2.41\\
1.00&4000&15.78&2.39\\
1.40&4000&43.79&3.03\\
1.60&4000&65.24&3.50\\
1.80&4000&91.06&3.98\\
2.00&4000&121.69&4.24\\
2.30&4000&174.01&6.69\\
2.50&4000&216.11&7.48\\
2.80&4000&294.00&6.12\\
0.80&8000&14.50&1.93\\
0.90&8000&18.96&2.72\\
1.20&8000&37.32&3.86\\
1.40&8000&58.22&4.69\\
1.60&8000&78.84&4.50\\
1.80&8000&107.73&6.17\\
2.00&8000&141.75&5.25\\
2.30&8000&203.00&5.72\\
2.50&8000&268.07&8.67\\
2.70&8000&308.87&12.26\\
\end{supertabular}
\end{center}

\begin{figure}
\includegraphics[width=0.8\linewidth]{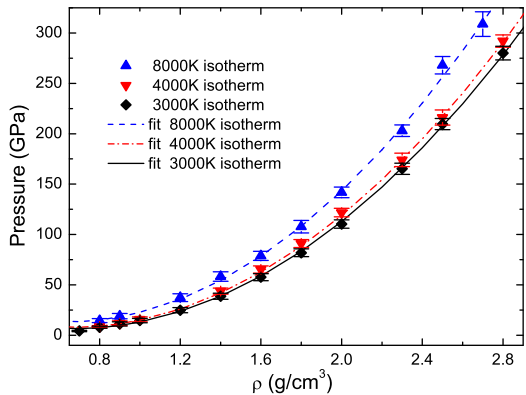}
\caption{(Color online) Pressure versus density along three isotherms ($T$=3000, 4000 and 8000 K). The lines are fitting
of the simulation results along isotherms.}
\end{figure}

The EOS has shown a systematic behavior in terms of density and
temperature. As has been reported previously, at temperatures
between 3000 and 4000 K along the principal Hugoniot, the
dissociation of methane occurs and nonmetal-metal phase transition
takes place simultaneously \cite{Li2011}. As a consequence, we
have studied the thermal EOS of fluid ethane along 3000, 4000 and
8000 K isotherms, as shown in Table I. The pressure $P$ consists
of two components $P_e$ and $P_i$. The gradient of total energy
calculated by VASP determines $P_e$, which represents the
contributions from ion-ion, ion-electron and electron-electron
interactions. Due to the classical movement of ions, the ionic
part $P_i$ is derived by the ideal gas expression. We thus have
$P=P_e+\rho_nK_BT$, where $\rho_n$ is the number density. The
simulated isotherm EOS for dense ethane is shown in Fig. 1. We
have fitted the simulated results for the pressure $P$ by
expansions in terms of density $\rho$ and temperature $T$ as
following $P={\sum}A_{ij}\rho^iT^j$. The coefficients $A_{ij}$ are
listed in Table II. This expansion can easily be applied in
hydrodynamic simulations for warm dense ethane in astrophysical
applications.

To clarify the structural transitions of ethane under extreme
conditions, we have calculated the pair-correlation functions
(PCFs), which represent the possibility of finding a particle at a
distance $r$ from a reference atom. The PCFs and the atomic
structure together with charge density distributions along 3000 K
isotherm are presented in Fig. 2 and 3, respectively. At the low
density $\rho$=1.60 g/cm$^3$, the peak in the C-H PCF
$g_\mathrm{C-H}(r)$ exists at around 1.09 {\AA} (Fig. 2a), which is
the equilibrium internuclear distance of the C-H bond in ethane
molecule. Meanwhile, the peak in the C-C PCF $g_\mathrm{C-C}(r)$
occurs at the equilibrium internuclear distance of C-C bond of
saturated hydrocarbon (1.54 {\AA}). Therefore, below the density of
1.60 g/cm$^3$, ethane remains in its ideal molecular configuration
without dissociation. From the structure of ethane at 1.60 g/cm$^3$
(Fig. 3a), we can straightly find that all of the ethane molecules
are not dissociated. At the density of 2.00 g/cm$^3$, we find a
significant reduction of the maximum of $g_\mathrm{C-H}(r)$ around
1.09 {\AA}, which indicates that the C-H bond breaks rapidly and
ethane molecules dissociate with the increase of the density. On the
other hand, around 0.75 {\AA} a shoulder appears in the H-H PCF at
this density, which implies the formation of molecular hydrogen. As
the density further increases to 2.3 g/cm$^3$, the maximum of the
PCF $g_\mathrm{C-H}(r)$ continues to reduce while the peak
amplitudes in $g_\mathrm{C-C}(r)$ increases, which indicates that
ethane further dissociates into some hydrocarbons. From Fig. 3c, we
can see that a new species of long alkane chains show up in the
atomic configuration, corresponding to the increase of the peak of
$g_\mathrm{C-C}(r)$. More hydrogen molecules forms at this density,
which corresponds to the larger value of $g_\mathrm{H-H}(r)$ around
0.75 {\AA}. Note that hydrogen molecules or atoms appear much more
in Fig. 3c than in Fig. 3b. However, the peak of $g_\mathrm{H-H}(r)$
around 0.75 {\AA} gets less apparent, and even vanishes at 2.8
g/cm$^3$. This may be because that (i) the peak of
$g_\mathrm{H-H}(r)$ around 0.75 {\AA} are covered by the widening of
the peak around the distance between the two hydrogen atoms bonding
with a carbon atom (i.e., more complicated alkane structures and
more hydrogen molecules and atoms form); and/or (ii) the formed
hydrogen molecules partially dissociated into hydrogen atoms.

\begin{table}[ptb]
\caption{Pressure (GPa) expansion coefficients $A_{ij}$ in terms of
density (g/cm$^3$) and temperature (K). }%
\begin{ruledtabular}
\renewcommand{\tabcolsep}{0.01pc}
\begin{tabular}{cccccc}
$j$&$A_{0j}$&$A_{1j}$&$A_{2j}$\\
\hline
0&27.3780&-73.3830&53.4723\\
1&0.0017&-0.0018&0.0020\\
\end{tabular}
\label{fit}
\end{ruledtabular}
\end{table}
From the reported of Li et al.\cite{Li2011}, it is known that above
the density 1.05 g/cm$^3$ along the Hugoniot curve (above the
pressure 25 GPa), methane begins dissociate and leads to the
appearance of ethane and some linear carbon chains. In our study on
the PCF of ethane, we find that ethane begins to dissociate at the
pressure 81 GPa, and hydrogen and alkane chains form. Compared with
the dissociation of methane at warm dense region \cite{Li2011}, both
of the hydrocarbons have the similar dissociation behavior with
increase of the density. The dissociated products of both alkane
species are hydrogen and alkane chains. Therefore, starting at
methane or ethane should result in basically the same final state
comprising alkane chains and hydrogen molecules and atoms at high
enough pressures and temperatures, except for the quantity of alkane
chains and molecular or atomic hydrogen. In addition, in the whole
dissociation process of ethane and methane simulated theoretically,
no diamond is formed, whereas it was experimentally reported that
the diamond is formed at pressure $P$=20 GPa and temperature
$T$=2000 K for methane\cite{Bene1999}. Richters $et$ $al.$
\cite{Rich} have employed a homogenous nucleation mechanism, which
is similar to the proposal of the direct graphite-to-diamond
transition \cite{Khal2011, Ghir2008}, to explain the discrepancy
between theory and experiment for methane.

\begin{figure}
\includegraphics[width=0.8\linewidth]{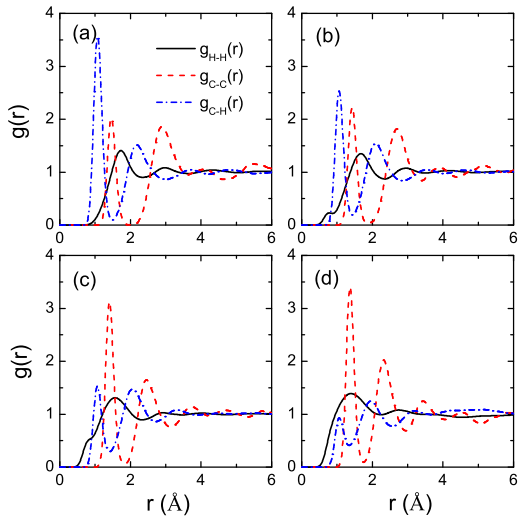}
\caption{(Color online) Pair-correlation functions for H-H (black
solid line), C-C (red dashed line), and C-H (blue dashed dot line)
for four densities of dense ethane along 3000 K isotherm. (a)
$\rho$=1.60 g/cm$^3$; (b) $\rho$=2.00 g/cm$^3$; (c) $\rho$=2.30
g/cm$^3$; (d) $\rho$=2.80 g/cm$^3$.}
\end{figure}
\begin{figure}
\includegraphics[width=0.8\linewidth]{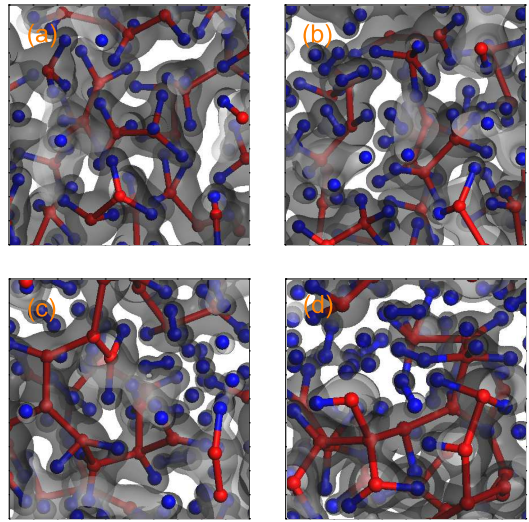}
\caption{(Color online) The atomic structure and charge density
distribution of dense ethane at four density along 3000 K isotherm.
carbon and hydrogen atoms are denoted by red and blue balls,
respectively. (a) $\rho$=1.60 g/cm$^3$; (b) $\rho$=2.00 g/cm$^3$;
(c) $\rho$=2.30 g/cm$^3$; (d) $\rho$=2.80 g/cm$^3$.}
\end{figure}
\subsection{Optical and electronic properties}

We have calculated the electronic conductivity and corresponding
reflectivity of fluid ethane along isotherms by Kubo-Greenwood
formula. The behavior of the frequency-dependent conductivity
$\sigma_1(\omega)$ at different densities along 3000 K isotherms is
shown in Fig. 4. It has been found that $\sigma_1(\omega)$ at these
four different densities exhibit a uniform feature that their peaks
all locate around 10.0 eV, which is correlated with the transitions
to the lowest excited states. With the increase of density along
isotherm, the main peak of the electrical conductivity
$\sigma_1(\omega)$ increases in amplitude and moves to low
frequency. Such change of the peak indicates another quantity, dc
conductivity, given as
$\sigma_{\rm{dc}}=\underset{\omega\rightarrow0}{\rm{lim}}
\sigma_{1}\left(\omega\right)$, increases with density. The dc
conductivity is extracted from dynamic conductivity and shown in the
inset of Fig. 4. The dc conductivity $\sigma_{\rm{dc}}$ at low
density along 3000 K isotherm is small, which corresponds to the
character of a semiconductor \cite{Nell1996,Nell1999}. With
increasing density, the dc conductivity $\sigma_{\rm{dc}}$ increases
rapidly. Above 2.30 g/cm$^3$ ($P$=165 GPa), a metal-like
conductivity is found. In order to analyze the effect of temperature
on the nonmetal-metal transition, we have also shown the dc
conductivity along higher temperature 4000 and 8000 K isotherms in
Fig. 4. At low density and high temperature the dc conductivity can
also reach a large value, which is a typical character of metallic
behavior.
\begin{figure}
\includegraphics[width=0.8\linewidth]{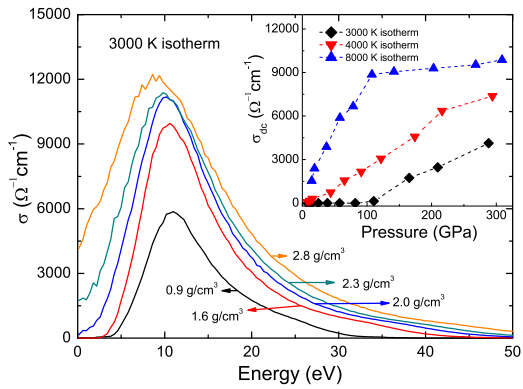}
\caption{(Color online) Optical conductivity spectra along the 3000
K isotherm. Data have been averaged over 15-20 uncorrelated MD
configuration. The dc conductivities as a function of density along
3000, 4000 and 8000 K isotherms have been shown in inset.}
\end{figure}

Mott \emph{et al.} \cite{Mott1990} have reported a decisive physical
assertion, which is used to define the metallicity of a disordered
system. When the characteristic mean free path of the valence
electrons exceeds the mean distance between the constituent atoms or
molecules, the states in disordered system with any high-temperature
will remain metallic or indeed attain metallic status. The
constituent atoms or molecules provide the carriers of the
electrical current. Applying Mott's criterion for the minimum
electrical conductivity for fluid metals, the minimum electrical
conductivity is about 1000 $\Omega^{-1}$cm$^{-1}$ for fluid
hydrogen, rubidium, caesium and mercury \cite{Nell2004}. The NMT has
recently been observed in the nominally nonmetallic chemical
elements hydrogen, nitrogen, and oxygen under extreme conditions
\cite{Nell2004,Nell2003}.
\begin{figure}
\includegraphics[width=0.95\linewidth]{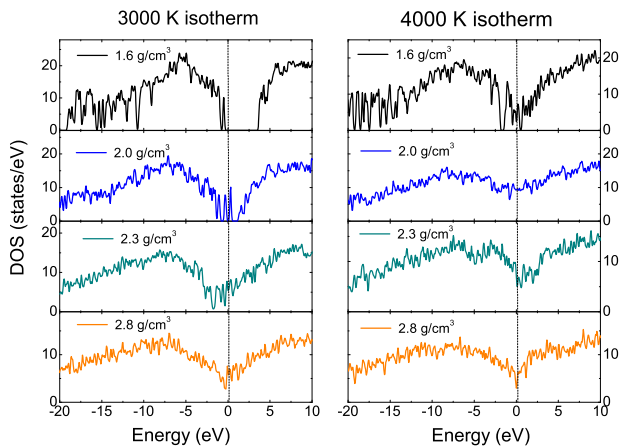}
\caption{(Color online) The density of state of fluid ethane at four
different density points along 3000 K isotherm. The Fermi energy is
set to be zero.}
\end{figure}

At high pressures, materials undergo dissociation bringing with some
new interesting properties. For instance, the band gap between the
valence and conduction bands decreases with pressure and then
disappears. The NMT has close relation with the molecular
dissociation in the shocked fluid. In the case of fluid hydrogen and
methane, QMD simulations have suggested that high-pressure NMT and
the dissociation of molecules with increasing pressure are closely
connected \cite{Li2011, Coll2001}. The dissociated atoms act as
dopants and progressively occupy the dense-fluid band gap. In order
to validate this physical picture, we have calculated the various
electron density of states at different densities along 3000 and
4000 K isotherms as shown in Fig. 5. For the low density 1.60
g/cm$^3$, a gap exits, locating near the Fermi energy where only
thermally activated electron transport occurs, which is exactly the
situation in semiconductors or nonmetals. With increasing the
density, the gap is gradually reduced, and finally disappears. At
density 2.30 g/cm$^3$, the electronic states fill up the gap so that
metal-like conductivity follows. In addition, Compared with the same
density along different isotherm, we can see the temperature effect
on the electronic structure. Such effect can also be visualized in
the dc conductivity $\sigma_{\rm{dc}}$-pressure plotted in the inset
of Fig. 4.

The optical reflectivity versus photon energy at different densities
along the 3000 K isotherm is shown in Fig. 6 (the main panel). With
the increase of density along the isotherm, the reflectivity
asymptotically approaches unity for zero photon energy. The
reflectivity versus density at fixed wavelength 404 nm
(corresponding to 3.075 eV) of fluid ethane along the 3000, 4000,
and 8000 K isotherms have been shown in the inset of Fig. 6. As is
seen, a measurable reflectivity arises from 0.10 to 0.49-0.59 at
typical wavelength of 404 nm spanning the visible spectrum, which is
related with the high-pressure NMT. Similar behavior in the optical
reflectance of fluid methane has been obtained by using QMD
simulations \cite{Li2011}. The results of the optical reflectivity
can be inspected through future experiments, and the reflectivity of
liquid deuterium has been executed in experiment \cite{Cell2000}.

\begin{figure}
\includegraphics[width=0.8\linewidth]{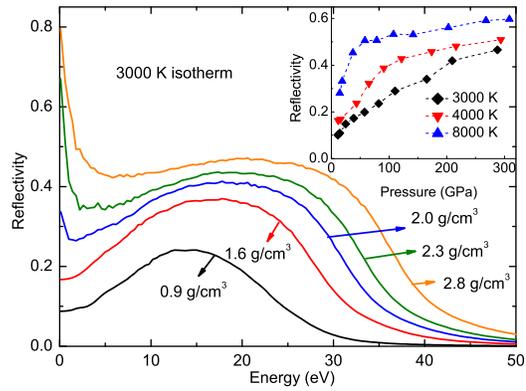}
\caption{(Color online) Variation of the frequency-dependent
reflectivity $r(\omega)$ along 3000 K isotherm. variation of the
reflectivity as a function of density at fixed wavelengths of 404 nm
along 3000, 4000 and 8000 K isotherms.}
\end{figure}

\section{CONCLUSIONS}
In summary, first-principles molecular-dynamics simulations based on
density-functional theory have been used to study the thermal EOS
and nonmetal-metal transition of fluid ethane under extreme
conditions. Systematic descriptions of pair-correlation function,
atomic structure, and the charge density distribution with various
densities along the isotherms are obtained to show the dissociation
process of molecular fluid ethane, which is associated with the
nonmetal-metal transition. Ethane dissociate with increasing the
density and temperature into molecular and/or atomic hydrogen and
some long alkane chains, which is similar to the case of methane.
However, no diamond-like carbon microstructure is observed for
ethane under extreme condition, the present result shows a different
high pressure behavior compared with previous experiments. The
nonmetal-metal transition has significant influence on the optical
properties of fluid ethane, which needs to be verified in dynamic
compression experiments. Our results are expected to be revealing
for the planetary models of Neptune.

\begin{acknowledgments}
This work was supported by NSFC under Grants No. 11005012,
No.11275032 and No. 51071032.
\end{acknowledgments}


\begin{thebibliography}{99}
\bibitem{Bene1999} L.R. Benedetti, J.H. Nguyen, W.A. Caldwell, H.J. Liu, M. Kruger, and R. Jeanloz, Science \textbf{286}, 100 (1999).

\bibitem{Stan2006} S. Stanley and J. Bloxham, Icarus \textbf{184}, 556 (2006).

\bibitem{Stan2004} S. Stanley and J. Bloxham, Nature \textbf{428}, 151 (2004).

\bibitem{Rich} D. Richters and T.D. K$\ddot{u}$hne, arXiv: 1206.4500v1.

\bibitem{Li2011} D. Li, P. Zhang, and J. Yan, Phys. Rev. B \textbf{84}, 184204 (2011).

\bibitem{Gao2010} G. Gao, A.R. Oganov, Y. Ma, H. Wang, P. Li, Y. Li, T. Lilaka, and G. Zou, J. Chem. Phys. \textbf{133}, 144508 (2010).


\bibitem{Vrie2007} J.de Vries, J.M. Hall, S.L. Simmons, M.J.A. Rickard, D.M.Kalitan, and E.L. Petersen, Combust. Flame \textbf{150}, 137 (2007).

\bibitem{Hida2000} Y. Hidaka, K. Sato, H. Hoshikawa, T. Nishimori, R. Takahashi, H. Tanaka, K. Inami, and N. Ito, Combust. Flame \textbf{120}, 245 (2000).

\bibitem{Frie1991} D.G. Friend, H. Ingham, and J.F. Ely, J. Phys. Chem. Ref. Data \textbf{20}, 275 (1991).


\bibitem{Wang2010} C. Wang and P. Zhang, J. Appl. Phys. \textbf{107}, 083502 (2010).

\bibitem{Cell2000} P.M. Celliers, G.W. Collins, L.B. Da Silva, D.M.
Gold, R. Cauble, R.J. Wallace, M.E. Foord, and B.A. Hammel, Phys.
Rev. Lett, \textbf{84}, 5564 (2000).

\bibitem{Leno2000} T.J. Lenosky, S.R. Bickham, J.D. Kress, and L.A. Collins, Phys. Rev. B \textbf{61}, 1 (2000).

\bibitem{Coll1995} L. Collins, I. Kwon, J. Kress, N. Troullier, and D. Lynch, Phys. Rev. E \textbf{52}, 6202 (1995).

\bibitem{Hohl1993} D. Hohl, V. Natoli, D.M. Ceperley, and R.M. Martin, Phys. Rev. Lett. \textbf{71}, 541 (1993).

\bibitem{Kres2010} J.D. Kress, J.S. Cohen, D.A. Horner, F. Lambert,
and L.A. Collins, Phys. Rev. E \textbf{82}, 036404 (2010).

\bibitem{Desj2002} M.P. Desjarlais, J.D. Kress, and L.A. Collins,
Phys. Rev. E \textbf{66}, 025401 (2002).

\bibitem{Kubo1957} R. Kubo, J. Phys. Soc. Jpn. \textbf{12}, 570 (1957).

\bibitem{Gree1958} D.A. Greenwood, Proc. Phys. Soc. London \textbf{71}, 585 (1958).

\bibitem{Maze2003} S. Mazevet, J.D. Kress, L.A. Collins, and P. Blottiau, Phys.
Rev. B \textbf{67}, 054201 (2003).

\bibitem{Maze2005} S. Mazevet, M.P. Desjarlais, L.A. Collins, J.D.
Kress, and N.H. Magee, Phys. Rev. E \textbf{71}, 016409 (2005).

\bibitem{Kres1993} G. Kresse and J. Hafner, Phys. Rev. B \textbf{47}, 558 (1993).

\bibitem{Kres1996} G. Kresse and J. Furthm$\ddot{u}$ller, Phys. Rev. B
\textbf{54}, 11169 (1996).

\bibitem{Perd1996} J.P. Perdew, K. Burke, and M. Ernzerhof, Phys. Rev. Lett. \textbf{77}, 3865 (1996).

\bibitem{Nose1984} S. Nos$\acute{e}$, J. Chem. Phys. \textbf{81}, 511 (1984).

\bibitem{Monk1976} H.J. Monkhorst and J.D. Pack, Phys. Rev. B \textbf{13}, 5188 (1976).

\bibitem{Khal2011} R.Z. Khaliullin, H. Esher, T.D. K$\ddot{u}$hne, J. Behler,
and M. Parrinello, Nature Mater. \textbf{10}, 693 (2011).

\bibitem{Ghir2008} L.M. Ghiringhelli, C. Valeriani, J.H. Los, E.J. Meijer, A. Fasolino,
and D. Frenkel, Mol. Phys. \textbf{106}, 2011 (2008).

\bibitem{Nell1996} W.J. Nellis, S.T. Weir, and A.C. Mitchell, Science \textbf{273}, 936 (1996).

\bibitem{Nell1999} W.J. Nellis, S.T. Weir, and A.C. Mitchell, Phys. Rev. B \textbf{59}, 3434 (1999).

\bibitem{Mott1990} N.F. Mott, \emph{Metal-Insulator Transitions, 2nd
edn.} (Taylor and Francis, London, 1990).

\bibitem{Nell2004} W.J. Nellis, J. Phys. Condens. Matter \textbf{16}, 923
(2004).

\bibitem{Nell2003} W.J. Nellis, R. Chan, P.P. Edwards, and R.
Winter, Z. Phys. Chem. \textbf{217}, 795 (2003).

\bibitem{Coll2001} L.A. Collins, S.R. Bickham, J.D. Kress, S.
Mazevet, T.J. Lenosky, N.J. Troullier, and W. Windl, Phys. Rev. B
\textbf{63}, 184110 (2001).




\end{thebibliography}
\end{document}